\documentclass[twocolumn]{article}

\usepackage[english]{babel}

\usepackage[letterpaper,top=2cm,bottom=2cm,left=3cm,right=3cm,marginparwidth=1.75cm]{geometry}

\usepackage{amsmath}
\usepackage{graphicx}
\usepackage{xurl}
\usepackage[colorlinks=true, allcolors=blue]{hyperref}
\usepackage[utf8]{inputenc}
\usepackage[T1]{fontenc}
\usepackage{enumitem}
\usepackage{xcolor}
\usepackage{colortbl}
\usepackage{array}
\usepackage{booktabs}
\usepackage{tcolorbox}
\usepackage{amssymb}
\usepackage{pythonhighlight}
\usepackage[frozencache,cachedir=minted-cache]{minted}
\usepackage{geometry}
\usepackage{enumitem}
\usepackage{wrapfig}
\usepackage{longtable}
\usepackage{authblk}
\usepackage{natbib}
\usepackage[subtle]{savetrees}

\setlist[itemize]{noitemsep, topsep=0pt, parsep=0pt, partopsep=0pt}
\setlist[enumerate]{noitemsep, topsep=0pt, parsep=0pt, partopsep=0pt}
\newcommand{\numpar}{1,000}

\title{\textbf{Private Map-Secure Reduce: Infrastructure for \\ Efficient AI Data Markets}}
\author[1]{Sameer Wagh}
\author[2]{Kenneth Stibler}
\author[1]{Shubham Gupta}
\author[1]{Lacey Strahm}
\author[1]{Irina Bejan}
\author[1]{Jiahao Chen}
\author[1]{Dave Buckley}
\author[1]{Ruchi Bhatia}
\author[1]{Jack Bandy}
\author[3]{Aayush Agarwal}
\author[1]{Andrew Trask}

\affil[1]{OpenMined Foundation}
\affil[2]{Reyvism Analytics}
\affil[3]{New York University}

\affil[ ]{\texttt {sameer@openmined.org, \{firstname\}@openmined.org}}
\affil[ ]{\texttt {ken@reyvism.com}, \texttt {aayush.agarwal@nyu.edu}}
\date{}

\begin{document}
\twocolumn[
\maketitle

\begin{abstract}
The modern AI data economy centralizes power, limits innovation, and misallocates value by extracting data without control, privacy, or fair compensation. We introduce Private Map-Secure Reduce (PMSR), a network-native paradigm that transforms data economics from extractive to participatory through cryptographically enforced markets. Extending MapReduce to decentralized settings, PMSR enables computation to move to the data, ensuring verifiable privacy, efficient price discovery, and incentive alignment. Demonstrations include large-scale recommender audits, privacy-preserving LLM ensembling (87.5\% MMLU accuracy across six models), and distributed analytics over hundreds of nodes. PMSR establishes a scalable, equitable, and privacy-guaranteed foundation for the next generation of AI data markets.
\end{abstract}
\vspace{4ex}]

\section{Introduction}\label{sec:intro}
The modern artificial intelligence ecosystem operates on a fundamentally inefficient data market. While computational capabilities have advanced exponentially, the economic architecture governing data (the essential input to AI systems) remains characterized by market failures that concentrate power, stifle innovation, and misallocate resources. The current paradigm, where data is extracted from individuals \citep{data50} and organizations \citep{cstd_data} with minimal control, privacy protection, or compensation, represents not merely an ethical concern but a profound economic inefficiency that undermines the sustainable potential of AI systems.

The central challenge lies in the unique characteristics of data that distinguish it from traditional goods: it is non-rivalrous in consumption, subject to significant network effects, and generates substantial externalities that are not captured in current pricing mechanisms \citep{acemoglu2022too}. These characteristics, combined with high transaction costs for negotiating data usage rights and enforcing privacy constraints, have led to market concentration among a few large platforms that can bypass these costs through ``free'' service models \citep{jones2020nonrivalry}.

This concentration creates multiple market failures. Information asymmetries prevent efficient price discovery, as data producers lack visibility into how their contributions are valued and used \citep{ichihashi2021economics}. Externalities---both positive spillovers from data contributions and negative privacy costs---remain unpriced, leading to systematic under-investment in data quality and over-extraction of sensitive information \citep{bergemann2019economics}. The resulting system exhibits characteristics of a ``tragedy of the commons,'' where the absence of clear property rights and governance mechanisms leads to degradation of our collective data, which carries serious risks for the health and sustainability of the entire AI ecosystem \citep{duch2017data}.

\begin{figure*}
    \centering
    \includegraphics[width=\linewidth]{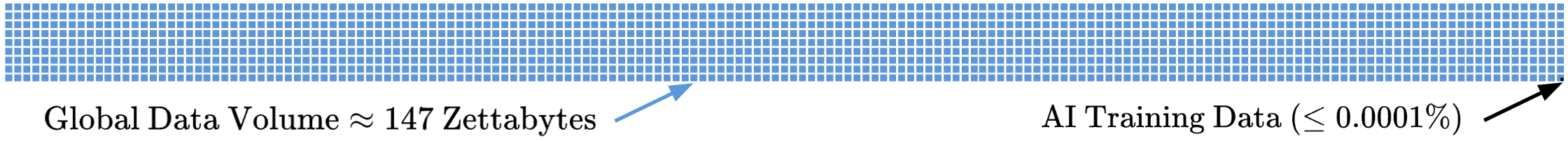}
    \caption{Less than 0.0001\% of global data is current used for AI training. Efficient data markets would allow a greater utilization of previously untapped data (GPT-5 estimate~\citep{gpt5data} and Global Data~\citep{statista_global_data_2024}}
    \label{fig:total_data}
\end{figure*}

Rather than assuming privacy as a good, we introduce Private Map-Secure Reduce (PMSR), a network-native paradigm to address these market failures through cryptographically-enforced data market infrastructure. PMSR inverts the traditional data pipeline by moving computation to data sources rather than extracting data to centralized systems. This architecture increases efficiency in the nascent market for data by reducing transaction costs, enabling price discovery, and internalizing privacy externalities.

The PMSR paradigm transforms data economics from extractive to participatory. Data contributors -- individuals, organizations, and institutions -- retain control over their information while participating in a cryptographically mediated market which compensates their contributions. This is achieved through a three-phase protocol: computation proposals that specify economic terms and privacy requirements, private map operations that execute computations locally while enforcing contributor-defined policies, and secure reduce operations that aggregate results without revealing individual contributions.

\section{The Market Failures of the Centralized AI Data Economy}

The centralized AI data economy is characterized by systematic market failures that embed inefficiency, suppress innovation, and concentrate power. These failures stem directly from the economic characteristics of data and the architectural of the AI ecosystem. In turn, understanding them is key to building more efficient and equitable systems~\citep{posner2018radical}.

\subsection{Information Asymmetries and Pricing Inefficiencies}

Information asymmetry arises when one party holds superior knowledge in a transaction. In data markets, this asymmetry is pervasive and systematic, creating fundamental barriers to efficient resource allocation. Without clear information on value and characteristics, markets undervalue data contributions, leaving producers undercompensated and incentives distorted across the ecosystem~\citep{arrieta2018should}.

Data producers, both individuals and organizations, rarely know how it is used, to whom it is transferred, or its value in model development~\citep{ghorbani2019data}. They are required to accept opaque terms of service that grant broad usage rights without an understanding of the value transferred. This information deficit prevents data producers from making informed decisions about data sharing and does not allow any negotiation of compensation.

While narrowly beneficial, this system means that data consumers face complementary uncertainties about data provenance, quality characteristics, and potential biases embedded in training datasets. This information gap can result in the development of flawed or biased models.

\subsection{Externalities and Unpriced Social Costs}

An externality represents a cost or benefit imposed on third parties who did not consent to incur that cost or benefit. The centralized AI data economy generates substantial negative externalities that are not reflected in market prices, leading to socially suboptimal outcomes.

Innovation externalities constitute a subtle but critical market failure. The concentration of data assets among a small number of platforms creates barriers to entry for new firms and researchers~\citep{martens}. This concentration prevents potential competitors from accessing the data necessary to develop competing AI systems, in turn reducing the AI ecosystem's efficiency and limiting the diversity of applications and approaches~\citep{khan2017market}.

Privacy harms constitute the most widely accepted negative externality in current data markets. The extraction and aggregation of personal data create substantial privacy risks including data breaches, unauthorized surveillance, and algorithmic manipulation~\citep{acquisti2016economics}. These costs are borne by data subjects and society broadly rather than by the platforms that profit from data extraction.

\subsection{Transaction Costs and Coordination Failures}

Transaction costs represent the costs of making an economic exchange, including search costs, negotiation costs, and enforcement costs. In data markets, these costs are prohibitively high, leading to market dominance by platforms that can bypass these costs through integrated service models~\citep{nagle}.

The costs of negotiating privacy-preserving data sharing agreements often exceed the expected benefits from collaboration. Legal frameworks for data sharing remain complex and fragmented, creating additional compliance costs. Technical standards for privacy-preserving computation are still emerging, requiring specialized expertise that many organizations lack~\citep{mitrea2023privacy}.

Even when data sharing would create positive sum outcomes, the costs of establishing trust, negotiating terms, and ensuring compliance prevent beneficial exchanges from occurring. This results in under-provision of collaborative research and reduced innovation in AI applications.

\section{The Private Map Secure Reduce for Efficient Data Markets }\label{sec:pmsr}

\begin{figure}[th]
    \centering
    \includegraphics[width=\linewidth]{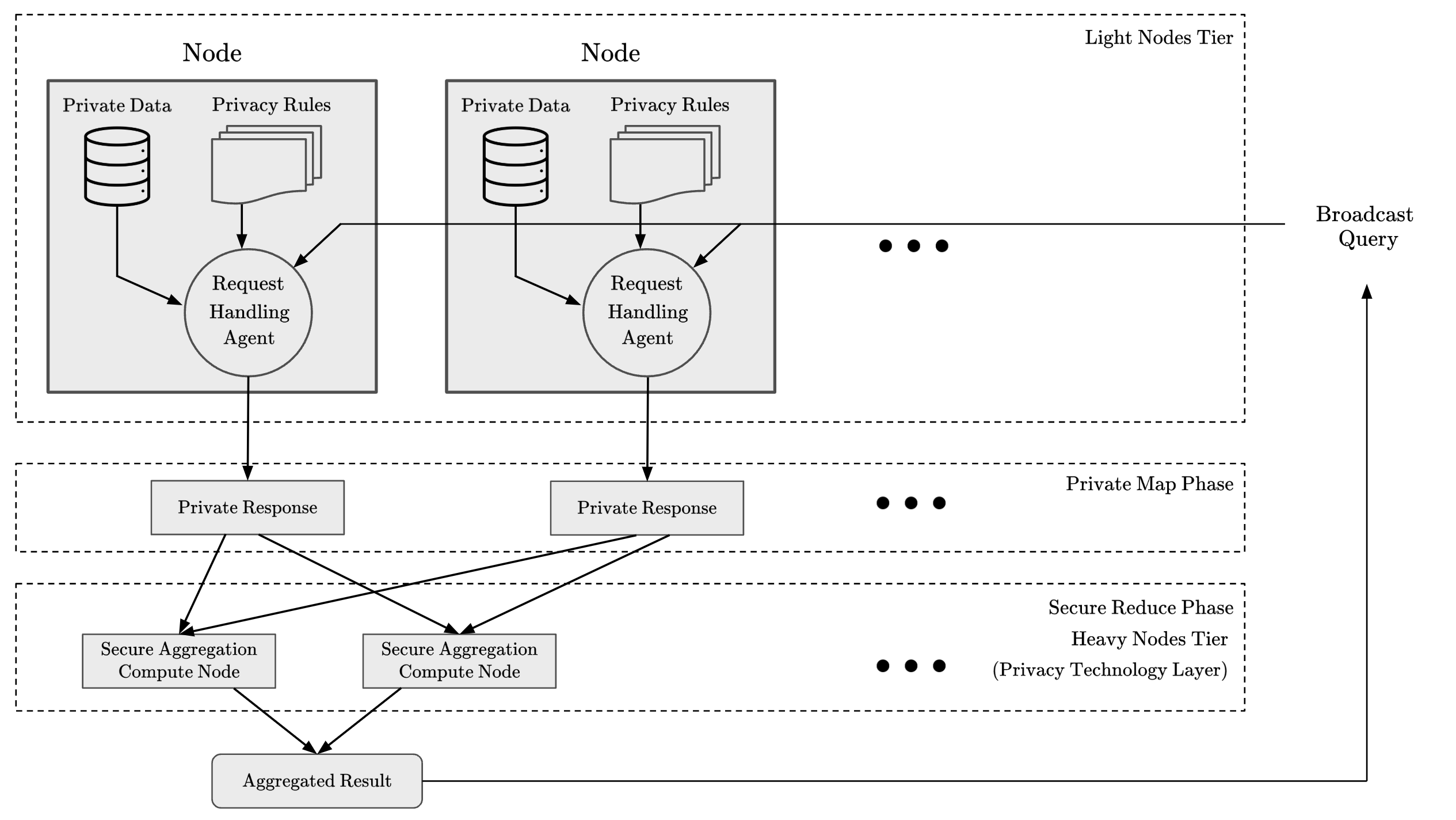}
    \caption{A schematic of network architecture. Light nodes, synonymous with data contributors, contain private data, their privacy rules, and a request handling agent that enforces the privacy rules on any computation. Heavy Nodes provide the necessary infrastructure to run privacy technologies for all users.}
    \label{fig:network_arch}
\end{figure}

Addressing the failures inherent in centralized data paradigms requires new market infrastructure. The Private Map-Secure Reduce (PMSR) paradigm is designed as this foundational infrastructure, providing technical alignment of data producer and consumer interests to address the root causes of market failures. PMSR thus takes the well known distributed computation MapReduce paradigm~\citep{mapreduce} and applies it at a network level (contrast with FL approaches~\citep{li2020federated}). Figure~\ref{fig:network_arch} shows a schematic of the network along with the lifecycle of a single computation (refer to Table~\ref{tab:computation_process} for a brief description of the phases).

\subsection{System Architecture}
\begin{figure}[t]
    \centering
    \includegraphics[width=\linewidth]{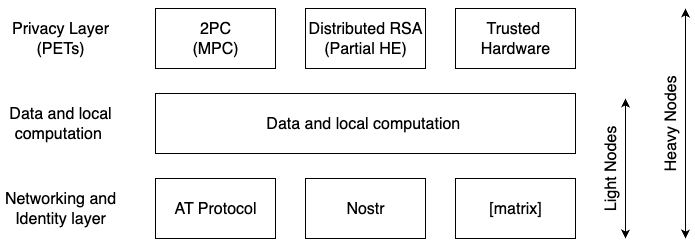}
    \caption{The architecture diagram for a Node.}
    \label{fig:nodearch}
\end{figure}

Each entity in this network is called a Node (cf~\ref{fig:nodearch}). A Node can be a server, cluster machine, an API or MCP server, or a personal computer or mobile device (via an application). The system makes a loose distinction between two types of Nodes: 
\begin{itemize}
    \item \textbf{Light Nodes}: Any entity that holds data and is willing to participate in computations (while maintaining a high degree of control over their data) can be a Light Node. 
    \item \textbf{Heavy Nodes}: Nodes can also choose to be be run on enterprise grade hardware and act as computation coordinators that orchestrate secure aggregation protocols without while protecting individual responses. 
\end{itemize}
The PMSR architecture is deliberately designed to balance decentralization with coordination efficiency. By separating Light Nodes (data owners) from Heavy Nodes (aggregation orchestrators), we enable flexible participation without imposing uniform infrastructure requirements. Light Nodes can range from personal devices to institutional databases, while Heavy Nodes operate as privacy-preserving coordinators, ensuring that no single party has unilateral access to raw data. This architecture naturally supports heterogeneous trust relationships, dynamic participant availability, and varying computational budgets -- conditions that mirror real-world deployment environments.

\subsection{Computation Proposal}
The Computation Proposal phase serves as the entry point for collaborative tasks in PMSR. Proposals can originate from any node -- individual researchers, institutions, or even autonomous agents -- and specify the computation logic, resource budget, privacy requirements, and intended recipients. The proposal is cryptographically signed to ensure authenticity and then broadcast to relevant portions of the network using a gossip-based protocol, reducing single points of failure. Nodes receiving the proposal locally evaluate its compatibility with their privacy policies before committing to participate. This decentralized matching mechanism ensures that participation is voluntary, policy-compliant, and economically motivated. Note that each computation flow assumes that the private map and secure reduce tasks are atomic to build higher-level functionality. Our implementation minimally specifies three parameters which define a computation object that encapsulates the distributed task in a structured and verifiable manner:
\begin{enumerate}
    \item Computation parameters such as a unique identifier, execution deadline, minimum participant threshold, economic budget, and optionally a set of targeted nodes.
    \item Private Map parameters including the function to be executed locally, with optional post-processing logic and an output schema describing the data format.
    \item Secure Reduce parameters defining the aggregation function applied to intermediate results, along with optional post-processing steps and a declared threat model to capture adversarial assumptions.
\end{enumerate}
Table~\ref{tab:computation_process} shows the three steps involved in the lifecycle of any computation over this network.

These standardized protocols eliminate the need for costly bilateral negotiations between data producers and consumers. The system provides a unified framework for specifying data usage rights, privacy constraints, and compensation mechanisms through its three-phase protocol: computation proposals, private map operations, and secure reduce operations. The cryptographic signing and gossip-based broadcasting protocol ensures authenticity while reducing single points of failure. Automated enforcement mechanisms embedded directly in the computational infrastructure eliminate the need for costly legal enforcement of data usage agreements. Privacy policies are enforced through local execution during the private map phase, where computation moves to data sources rather than requiring data transfer. This architectural choice ensures that data usage rights are enforced cryptographically rather than contractually.

\begin{table*}[tbp]
\centering
\resizebox{\textwidth}{!}{%
\renewcommand{\arraystretch}{1}
\setlength{\tabcolsep}{8pt}
\begin{tabular}{>{\columncolor{blue!8}}c >{\columncolor{red!12}}p{152mm}}
\toprule
\rowcolor{blue!15}
\textbf{Step} & \textbf{Description} \\
\midrule
\midrule
\textbf{1} & 
\textbf{Computation Proposal}
\newline
Any Node in the system can propose a computation. A computation specifies target participants, the program/analysis, secure computation threat model, output schema, economic budget etc. The computation is then asynchronously broadcast to the appropriate Nodes in the network.\\
\midrule
\textbf{2} & 
\textbf{Private Response (Private Map)}
\newline
Local systems that enable interaction with the network. Each Node responds to the computation by respecting the Nodes' personalized privacy policies (in other words each system has complete control over it's interaction with the network). Output adheres to the specified output schema.\\
\midrule
\textbf{3} & 
\textbf{Secure Aggregation (Secure Reduce)}
\newline
Selected nodes perform secure aggregation protocol (a modular interface for privacy technologies such as MPC, HE, TEE etc.). Secure aggregation will ensure that that individual Node responses are never revealed; only aggregate responses are outputs of the computation if the participant thresholds are met.\\
\bottomrule
\end{tabular}%
}
\caption{\textbf{3-Step Distributed Computation Process.}}
\label{tab:computation_process}
\end{table*}

\subsection{Private Map}
In the Private Map phase, computation moves to where the data resides. Each participating Light Node executes the map function locally, transforming raw data into intermediate, privacy-compliant representations. Privacy policies -- which can be declarative (e.g., ``no raw PII leaves the node'') or procedural (e.g., ``require manual approval for sensitive queries'') -- are enforced in situ~\citep{cheu2019distributed}. PMSR supports both static policy definitions and adaptive policy agents that can interpret and respond to novel computation requests. The output schema is validated before transmission to ensure data normalization across heterogenous Nodes.

\begin{minted}[fontsize=\small,baselinestretch=0.9]{python}
# Node requires trusted org, blocks sensitive 
# data, enforces participant thresholds, and 
# allows aggregated health stats.
health_privacy_agent = RequestHandler(
  llm_model='gpt-4o',  # claude-4o etc.
  privacy_policy=(PrivacyPolicy()
    .require("@trusted-domain.org email")
    .block("PII in output")
    .require("100+ participants")
    .allow("aggregated functions"))
)

# A self-hosted model that manually approves
# all queries and requires a dishonest 
# majority MPC computation for aggregation.
research_privacy_agent = RequestHandler(
    llm_model='llama-3',
    privacy_policy=(PrivacyPolicy()
        .require("Manual approval")
        .require("Dishonest majority MPC"))
)
\end{minted}

\subsection{Secure Reduce and Cryptographic Guarantees}
Secure Reduce phase aggregates intermediate results without revealing individual contributions, even to Heavy Nodes. PMSR integrates modular secure computation protocols -- such as multi-party computation (MPC)~\citep{bonawitz2017federated, elsa, corrigan2017prio}, homomorphic encryption (HE)~\citep{FHE, gentry2009fully}, or trusted execution environments (TEE)~\citep{zheng2017opaque, tee} -- allowing computations to tailor their privacy guarantees to their threat model. Below are a few examples of secure reduce computation models. These are just a few examples of how secure computation frameworks can be applied; the modular nature of the infrastructure ensures that other PETs solutions, with threat models appropriate to the use-case in consideration, can also be seamlessly integrated at the Heavy Nodes:

\noindent\textbf{3PC Secure Computation.} One approach is to use three-party secure computation protocols~\citep{SP:ABFLLN17, FLNW17,sameerthesis}\nocite{securenn,falcon,elsa,dpcrypto,pika}. In a 3-party semi-honest MPC protocol with modulus $2^{64}$, each Light Node secret-shares every value if it's local result $x$ into $(x_1, x_2, x_3)$ such that:
\begin{equation}
x \equiv \sum_{i=1}^3 x_i \pmod{2^{64}}
\end{equation}

Each share is sent to a distinct Heavy Node $H_i$. The Heavy Nodes jointly compute the reduction function on these shares without ever reconstructing the individual $x$ values. Only the final aggregate, if it passes anonymity thresholds, is revealed. This mechanism enforces privacy by design rather than by policy, making leakage due to misconfiguration or malice substantially harder. MPC can be used to run arbitrary computations include LLMs as shown in the recent work Orca~\citep{orca}.

\noindent\textbf{HE based aggregation.} Distributed RSA~\citep{BFdistributedRSA} is an algorithm where a set of parties jointly hold a RSA key $N = pq$ where no party knows the prime factorization. Similarly, the parties know a public exponent $e$ and share a private exponent $d$ (each party holds a share of $d$) such that $d\cdot e \equiv 1 \pmod{\phi(N)}$. Then secure aggregation works as follows: each Light Node encrypts and shares every value $x$ to the Heavy Nodes as $\mathsf{Enc}_N(x)$. Since RSA is additively homomorphic, different values can be aggregated together without decrypting. 
\begin{equation}
    \mathsf{Enc}_N(x_A) \cdot \mathsf{Enc}_N(x_B) = \mathsf{Enc}_N(x_A+x_B)
\end{equation}
A single round of distributed decryption can be run using the distributed decryption key to reveal the aggregated answer. Other ways include one party generating the keys and another party running the HE computation using a framework like THOR~\citep{miran_thor}.

\noindent\textbf{TEE Based Aggregation.} 
Trusted Execution Environments (TEEs) such as Intel SGX or AMD SEV provide a hardware-based mechanism for isolating computation inside an enclave that cannot be tampered with by the host operating system~\cite{tee}. In PMSR, Heavy Nodes can act as enclaves that execute the reduction function on encrypted intermediate values. Light Nodes encrypt their local outputs and transmit them to the enclave, which then decrypts and processes them entirely within the protected environment. Remote attestation protocols allow Light Nodes to verify that the correct computation code is running inside the enclave before contributing their data. Only the final aggregate result is released from the enclave, ensuring that raw intermediate values remain inaccessible even to the host of the Heavy Node. While TEEs introduce hardware trust assumptions and potential side-channel risks, they provide strong practical guarantees with relatively low computational overhead compared to fully cryptographic protocols.

For data consumers the system's modular interface with multiple PETS enables secure aggregation which enables desirable analytics and modeling activities. This internalization of positive externalities creates appropriate incentives for ongoing high-quality data contribution.

\subsection{Empirical Validation of Viability}

Our evaluation demonstrates PMSR's viability through evaluations over different scenarios and application domains. The system enables previously impossible collaborations: external researchers conducting algorithmic audits without accessing raw data, distributed AI inference achieving performance parity with centralized systems, and large scale collaborative statistics while maintaining individual privacy.

These results suggest that PMSR provides not only technical feasibility but economic viability for a more efficient and equitable AI data economy~\citep{blockchainIncentives}. The system's ability to enable secure collaboration in contexts that were previously too sensitive to share demonstrates its potential to address the coordination failures that characterize current data markets.

The broad adoption necessary for network effects is incentivized through increasing returns for participants: more contributors improve model quality, which attracts more data consumers, who provide greater compensation to contributors. This creates a positive feedback loop that aligns individual incentives with collective benefits.

\section{Evaluation}
We evaluate PMSR extensively with a series of real-world deployments that span diverse computational tasks and participation scales:

\begin{figure*}[h]
    \centering
    \includegraphics[width=\linewidth]{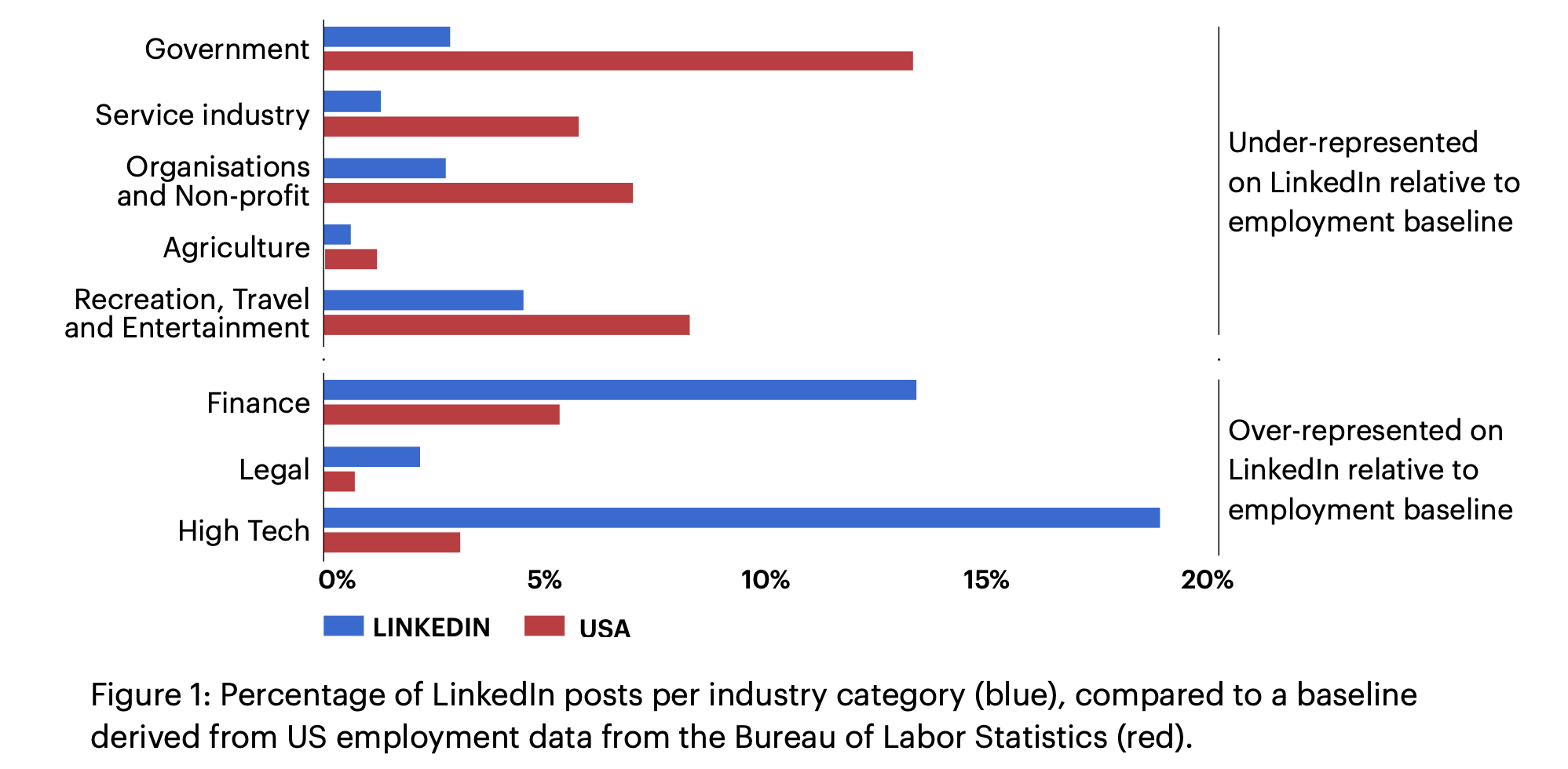}
    \caption{Industry representation on LinkedIn versus U.S. employment baseline (from Bureau of Labor Statistics). High-tech, legal, and finance sectors show $2$-$3\times$ over-representation, while service and government sectors are under-represented by similar magnitudes.}
    \label{fig:linkedin_results}
\end{figure*}

\begin{enumerate}
    \item The first experiment shows the benefits and feasibility of PMSR through a real-world deployment enabling third-party algorithmic audits of production recommender systems at LinkedIn and Dailymotion. Four independent researchers successfully analyzed over 80 million data points across both platforms without ever accessing raw user data, demonstrating PMSR's practical viability for privacy-preserving computation. We present the LinkedIn results in Section~\ref{sec:linkedin} and defer the Dailymotion reults to Appendix~\ref{app:linked_daily}.
    \item The distributed AI inference experiment demonstrates PMSR's capability to enable collaborative model ensembling across organizational boundaries without sharing proprietary model weights or raw predictions. Six different LLMs (hosted by different hypothetical organizations) collaborate through PMSR to achieve 87.5\% accuracy on the MMLU benchmark -- a 3\% improvement over the best individual model (gpt-4o at 84.8\%). This validates PMSR's potential to unlock the economic value of model diversity while preserving intellectual property (cf Section~\ref{sec:sota}).
    \item The large-scale distributed statistics experiment validates PMSR's scalability through privacy-preserving aggregation of sensitive health data across \numpar{} participants. Using 3-party secret sharing among 50 Heavy Nodes, the system computes population-level statistics while ensuring no party can reconstruct individual contributions (cf Section~\ref{sec:llm}).
    \item Finally, we explore the use of a Mock/Real paradigm and it's effectiveness in resolving the fundamental tension in the competing demands of data exploration and privacy protection (cf Section~\ref{sec:mock}).
\end{enumerate}

\subsection{LinkedIn Audit: Algorithmic Bias in Professional Networks}\label{sec:linkedin}

\noindent\textbf{Experimental Setup.} LinkedIn provided impression data from 70 million feed ranking events across 18 industry categories. Rather than sharing raw data, PMSR's Private Map phase allowed Light Nodes (the platforms) (1) locally execute audit queries (2) enhance response privacy using differential privacy~\citep{dwork2014algorithmic}, implemented through OpenDP integration~\cite{opendp_library}. The auditors operated remotely from distributed locations (3 in the US, 1 in the UK) to run the analysis. 

\textbf{Private Map and Secure Reduce Implementation.} Each Light Node executed local queries against feed impression logs, extracting features including industry category, educational attainment, and job advertisement exposure. The light node provided mock data as a sight-aid to experiment and close in on the analysis code. The study involved questions around bias in viewing activity across industries and the effect of the choice of recommendation algorithms for job ad serving. The privacy policy automatically enforced LinkedIn's differential privacy budget for each request. Secure Reduce, in this study, was done using a single Heavy node since the access control (provided by the network node identity) and the differential privacy guarantees were sufficient for this use case (contrast with DailyMotion in Appendix~\ref{app:linked_daily}).

\textbf{Results and Impact.} 
Third-party algorithmic audits are crucial for AI transparency, enabling the identification of risks such as unfair bias, intellectual property violations, and the spread of disinformation or other harmful content~\citep{auditml}. PMSR makes such audits feasible at scale through privacy guarantees that give platforms confidence to share production data. LinkedIn (and Dailymotion) made real, anonymized data available to external researchers precisely because the system ensured no auditor would observe individual-level records and custom privacy mechanisms could be enforced (differential privacy). Figure~\ref{fig:linkedin_results} demonstrates the value of this approach through a study that reveals significant demographic skew in platform participation. Knowledge work sectors (technology, finance, legal) were over-represented by factors of $2$-$3\times$ relative to Bureau of Labor Statistics employment baselines, while manual labor sectors showed inverse patterns. These findings validate PMSR's ability to uncover subtle algorithmic biases while maintaining strict privacy guarantees: statistical patterns emerged clearly yet individual participant privacy remained protected throughout the analysis.

\subsection{Distributed AI Inference: Collaborative Model Ensembling}\label{sec:sota}

\noindent\textbf{Experimental Setup.} To demonstrate PMSR's applicability beyond data auditing, we evaluate its capability for distributed AI inference where multiple organizations collaborate without sharing model weights or raw predictions. Six state-of-the-art language models-gpt-4o, grok-2, deepseek, phi-4, mixtral-8x22b, and llama-3.2-participated in collaborative inference on the MMLU benchmark spanning 57 diverse subjects. Each model, hosted by a different Light Node representing distinct organizations, maintained complete control over its proprietary weights while contributing to ensemble predictions. The experiment required a minimum of 5 participating models to ensure robustness, with the computation proposal specifying Gradient-Aware Calibrated (GaC) ensembling~\cite{yao2024gac} as the aggregation protocol.

\begin{figure*}[th]
    \centering
    \includegraphics[width=\linewidth]{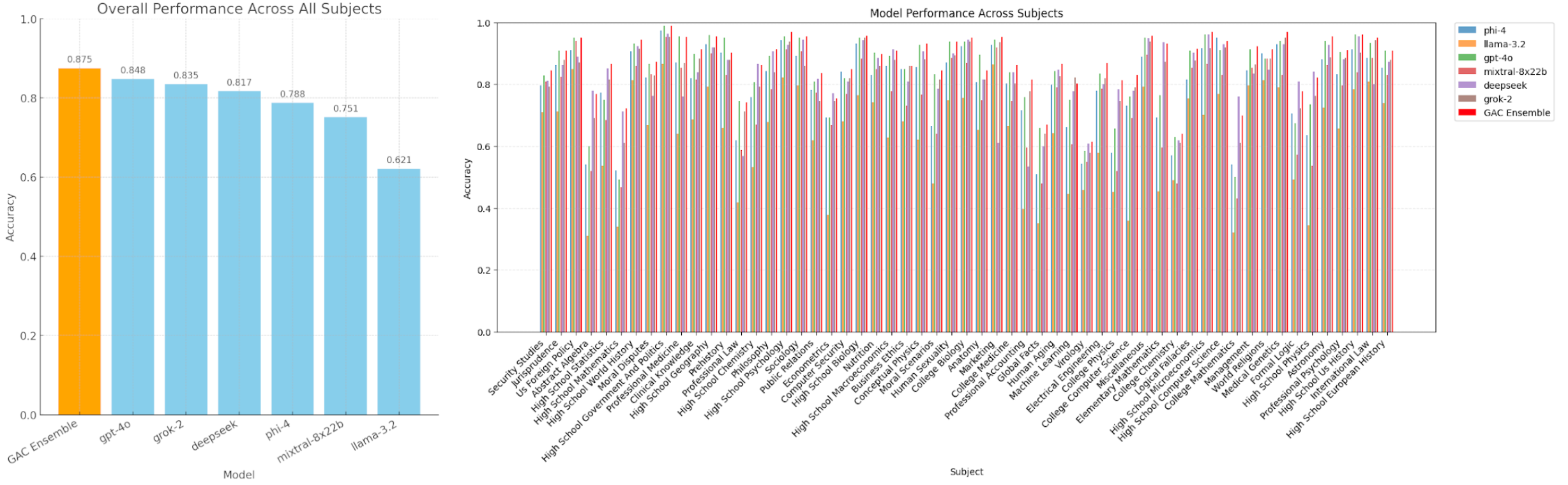}
    \caption{Performance comparison of individual models versus GaC ensemble on MMLU. The ensemble (87.5\%) outperforms the best individual model (gpt-4o at 84.8\%) by 3.15\%. Right panel shows per-subject accuracy differences, with ensemble improvements in 42 of 57 subjects.}
    \label{fig:ensemble}
\end{figure*}

\textbf{Private Map and Secure Reduce Implementation.} In the Private Map phase, each Light Node computed log probabilities for MMLU questions locally, applying model-specific normalization without revealing raw outputs. The nodes quantized output embeddings and secret-shared them among Heavy Nodes before aggregation. Privacy policies enforced a minimum participation threshold of 5 models, ensuring no single model's predictions could be reverse-engineered. The Secure Reduce phase implemented the GaC protocol: Heavy Nodes securely computed weighted averages of log probabilities using Bayesian-optimized weights (gpt-4o: 0.349, grok-2: 0.303, deepseek: 0.347) without reconstructing individual predictions. This architecture preserved each model's intellectual property while enabling collaborative intelligence.

\textbf{Results and Impact.} Figure~\ref{fig:ensemble} demonstrates that distributed inference through PMSR achieves meaningful performance gains while preserving model sovereignty. The GaC ensemble reached 87.5\% accuracy, surpassing the best individual model by 3.15 percentage points. Subject-wise analysis revealed heterogeneous improvements: the ensemble particularly excelled in interdisciplinary topics requiring diverse knowledge (e.g., High School European History: +5.2pp, College Medicine: +4.8pp) while showing minimal gains in highly specialized domains where individual models already performed well. These results validate PMSR's economic proposition: organizations can monetize their model investments through collaborative inference markets without surrendering competitive advantages. The gap between ensemble performance (87.5\%) and theoretical maximum (95.9\%) suggests substantial untapped value in broader model collaboration, achievable only through privacy-preserving infrastructure like PMSR.

\subsection{Large Scale Distributed Statistics}\label{sec:llm}
\noindent\textbf{Experimental Setup.} To evaluate PMSR's scalability for population-level statistics, we deployed a network of \numpar{} Nodes representing individual users contributing sensitive health data, specifically annual sleep score averages. The computation infrastructure utilized 50 Heavy Nodes (remaining Light Nodes) configured for 3-party secret sharing, ensuring that secure aggregation could proceed even with up to node failures upto the  minimum threshold of the computation participants. Each participant contributed their locally-computed annual average without revealing individual measurements, testing PMSR's feasibility to handle privacy-sensitive personal analytics over a small network.

\textbf{Private Map and Secure Reduce Implementation.} In the Private Map phase, each Light Node computed a one-year rolling average of daily sleep scores locally. Rather than transmitting raw averages, each Light Node split its value into three secret shares using Shamir's secret sharing scheme~\citep{SS} with a threshold of 2. These shares were distributed across the selected triplet of Heavy Nodes, ensuring no single Heavy Node or even pairs of Heavy Nodes could reconstruct any participant's data. The Secure Reduce phase leveraged the additive homomorphic properties of secret sharing: Heavy Nodes computed the population mean by summing their shares and only reconstructing the final aggregate. The protocol enforced a minimum participation threshold of 500 nodes to ensure statistical privacy through aggregation, automatically aborting if insufficient participants joined.

\begin{figure*}[t]
    \centering
    \includegraphics[width=\linewidth]{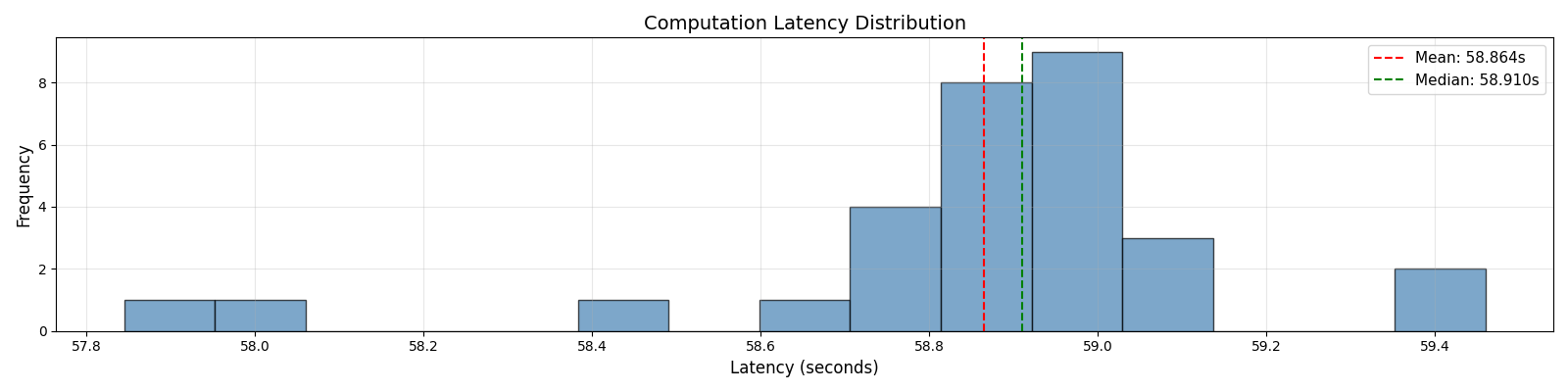}
    \caption{Experiment over a Network of \numpar{} Nodes. PMSR allows aggregate computation of the sleep scores of individuals without any participant revealing their individual sleep score to any entity in the Network.}
    \label{fig:latency}
\end{figure*}

\textbf{Results and Impact.} Figure~\ref{fig:latency} demonstrates PMSR's feasibility over a large distributed computation:  The computation deadlines were set to a 10 second deadline and the system achieved mean computation latency of 58.864 seconds (median: 58.910s) for secure aggregation across \numpar{} Nodes, with remarkably low variance. The tight latency distribution validates PMSR's architectural choices: the separation of local computation (Private Map) from heavy cryptographic operations (Secure Reduce) enables complexity proportional to the number of Heavy Nodes involved in the computation. This performance profile makes PMSR viable for real-world applications requiring frequent population statistics: public health monitoring, distributed clinical trials, and privacy-preserving social science research. This provides a new avenue for privacy-first computing of large-scale population statistics over sensitive data.

\subsection{Mock Real Paradigm}\label{sec:mock}
The fundamental tension in privacy-preserving analytics lies in the competing demands of data exploration and privacy protection. Data scientists require iterative experimentation to develop effective analyses, yet each interaction with sensitive data incurs privacy costs and risks exposure. Traditional approaches force analysts to work blindly, submitting queries to black-box systems without visibility into data structure or intermediate results. This leads to inefficient use of privacy budgets, high query failure rates, and ultimately, underutilization of valuable data assets.

We introduce the Mock/Real paradigm, a dual-node architecture that resolves this tension through separation of concerns. Each data contributor operates two nodes: a mock node containing representative samples (few samples of real data, or synthetic data, or historical snapshots) with relaxed privacy constraints, and a real node hosting complete data with full privacy enforcement. This separation enables analysts to iterate freely on mock data-debugging queries, testing joins, and optimizing algorithms-before executing refined computations on real data.

In the LinkedIn study, without access to mock data, it would have been impossible to successfully run queries. Despite the mock data, we met unforeseen issues when transitioning from using the local mock data to the remote production data. This was primarily due to the mock data being smaller in size, and having a `cleaner' distribution that was Gaussian with few outliers (cf. the production data which was non-Gaussian with long tails and many outliers). This required an iterative process to learn about the data from subsequent requests to construct the analysis, as the mock data properties proved to be sometimes misleading~\cite{christchurch2024}.

\section{Discussion and Related Works}\label{sec:discussion}
PMSR aligns with the tightening regulatory environment by embedding privacy guarantees directly in computation and by providing solid foundation for data visitation systems. Unlike federated learning~\citep{Kairouz_fl}, PMSR supports arbitrary computations beyond model training, with incentives for both Light (data holders) and Heavy (aggregator) nodes. It bridges policy and practice: data remains locally governed, yet aggregate insights flow, enabling AI to scale into the orders-of-magnitude larger private data universe.

\noindent\textbf{State of Digital Privacy.} While regulations such as GDPR nominally grant individuals control over their data, practical enforcement remains weak and often limited to data custody rather than usage~\citep{hoofnagle2019european}. In reality, privacy breaches frequently occur after data has been voluntarily shared with a ``trusted'' entity~\citep{solove2021breach}. PMSR addresses this gap by allowing participants to inspect the computation body itself, enabling proactive detection of unintended or unauthorized use before data leaves the local environment, similar to verifiable computation approaches~\citep{gennaro2010non}.

\noindent\textbf{Selecting Heavy Nodes.} The integrity of Heavy Nodes is central to PMSR's trust model. Lessons from decentralized networks such as Tor~\citep{dingledine2004tor} suggest that distributed trust, combined with public reputation mechanisms, can mitigate risks. Byzantine fault-tolerant consensus protocols~\citep{castro1999practical} and blockchain-based reputation systems~\citep{dorri2017blockchain} provide theoretical foundations for our approach. We envision community-driven reputation scoring, cryptographic attestations of correct protocol execution, and periodic rotation of Heavy Node roles to limit collusion opportunities.

\noindent\textbf{Incentives and Sustainability.} Sustaining long-term participation requires clear, equitable incentives. Game-theoretic approaches to incentive design in distributed systems~\citep{roughgarden2010algorithmic} inform our mechanism design. For Heavy Nodes, rewards may come in the form of direct compensation, access to aggregated results, or enhanced network reputation. Light Nodes can be compensated directly by computation sponsors, receive proportional credit for contributions to public datasets~\citep{jia2019towards}, or gain privileged access to network services. By aligning economic incentives with privacy preservation, PMSR makes privacy a value-add rather than a compliance cost. In a similar spirit, blockchain-based incentive mechanisms for federated learning systems are explored in~\citep{blockchainIncentives}.

\noindent\textbf{Relation to Existing Tools.} Several projects, such as Pol.is, RadicalxChange's quadratic voting demos, and FedML~\citep{he2020fedml}, aim to facilitate collective decision-making or decentralized model training. DAP~\cite{dap-16} has a very similar goal to this work but is more restrictive in terms of the aggregation models it supports. Other frameworks like FATE~\citep{liu2021fate}, PySyft~\citep{ryffel2018generic}, FedMapReduce~\cite{fedmapreduce} provide federated learning infrastructure but lack privacy or economic incentive layers. Recent work from Anthropic, called Clio~\cite{clio}, performs user log analysis for Claude (over a million conversations) and can be seen as an in-house instantiation of the same abstractions proposed by PMSR (mapping to Facets and Initial clusters and minimum threshold enforcing would comprise private mapping whereas the clustering is done in the clear but can be enhanced using secure reduce). PMSR thus differs in scope and design: it is a network-native abstraction layer for arbitrary secure computations, capable of scaling across diverse domains and integrating multiple privacy-preserving technologies under a unified programming model.

\noindent\textbf{From Policy to Architecture.} Unlike centralized cloud computing or traditional federated learning, PMSR's privacy enforcement is embedded directly into the computational substrate. This shifts the locus of trust from institutional promises to cryptographic guarantees~\citep{goldwasser2019cryptographic}, creating a by-default privacy layer that is resilient to policy drift or administrative overreach. This approach aligns with the principles of privacy by design~\citep{cavoukian2009privacy} but implements them through cryptographic enforcement rather than organizational compliance. There is an emerging focus on having social media move to decentralized systems and PMSR can be seen as a social media for data use~\citep{newpublic_blacksky}.

\noindent\textbf{Scalability Considerations.} With $N$ total nodes and $n$ Heavy Nodes, PMSR's preprocessing cost scales as $O(n)$ for each Heavy Node's communication and storage. Our prototype shows that this is feasible for $n \ll N$ even at global scale, provided that network bandwidth and cryptographic preprocessing are efficiently managed. Advances in scalable secure computation~\citep{mohassel2018aby3} and communication-efficient protocols~\citep{damgard2012multiparty} support this scaling behavior. Heavy Nodes in our experiments automatically accepted all valid computation requests, ensuring high throughput without centralized bottlenecks.

\section{Conclusion}
Unlocking the next frontier of AI requires rethinking incentives, privacy, and architecture. PMSR offers a principled path: computation moves to the data, sovereignty is preserved, and participation is economically aligned. By doing so, PMSR transforms private data from locked silos into a foundation for lawful, global-scale, data-driven intelligence.

\nocite{langley00}

\bibliography{example_paper}

\begin{thebibliography}{64}
\providecommand{\natexlab}[1]{#1}
\providecommand{\url}[1]{\texttt{#1}}
\expandafter\ifx\csname urlstyle\endcsname\relax
  \providecommand{\doi}[1]{doi: #1}\else
  \providecommand{\doi}{doi: \begingroup \urlstyle{rm}\Url}\fi

\bibitem[Acemoglu et~al.(2022)Acemoglu, Makhdoumi, Malekian, and Ozdaglar]{acemoglu2022too}
Acemoglu, D., Makhdoumi, A., Malekian, A., and Ozdaglar, A.
\newblock Too much data: Prices and inefficiencies in data markets.
\newblock \emph{American Economic Journal: Microeconomics}, 14\penalty0 (4):\penalty0 218--256, 2022.

\bibitem[Acquisti et~al.(2016)Acquisti, Taylor, and Wagman]{acquisti2016economics}
Acquisti, A., Taylor, C., and Wagman, L.
\newblock The economics of privacy.
\newblock \emph{Journal of Economic Literature}, 54\penalty0 (2):\penalty0 442--492, 2016.

\bibitem[AMD(2020)]{tee}
AMD.
\newblock {AMD SEV-SNP: Strengthening VM Isolation with Integrity Protection and More}.
\newblock White Paper, 2020.
\newblock URL \url{https://www.amd.com/content/dam/amd/en/documents/epyc-business-docs/white-papers/SEV-SNP-strengthening-vm-isolation-with-integrity-protection-and-more.pdf}.

\bibitem[Araki et~al.(2017)Araki, Barak, Furukawa, Lichter, Lindell, Nof, Ohara, Watzman, and Weinstein]{SP:ABFLLN17}
Araki, T., Barak, A., Furukawa, J., Lichter, T., Lindell, Y., Nof, A., Ohara, K., Watzman, A., and Weinstein, O.
\newblock Optimized honest-majority mpc for malicious adversaries -- breaking the 1 billion-gate per second barrier.
\newblock In \emph{IEEE Symposium on Security and Privacy (S\&P)}, 2017.

\bibitem[Arrieta-Ibarra et~al.(2018)Arrieta-Ibarra, Goff, Jim{\'e}nez-Hern{\'a}ndez, Lanier, and Weyl]{arrieta2018should}
Arrieta-Ibarra, I., Goff, L., Jim{\'e}nez-Hern{\'a}ndez, D., Lanier, J., and Weyl, E.~G.
\newblock Should we treat data as labor? moving beyond "free".
\newblock \emph{American Economic Review: Papers \& Proceedings}, 108:\penalty0 38--42, 2018.

\bibitem[Bergemann \& Bonatti(2019)Bergemann and Bonatti]{bergemann2019economics}
Bergemann, D. and Bonatti, A.
\newblock The economics of social data: An introduction.
\newblock \emph{Annual Review of Economics}, 11:\penalty0 439--466, 2019.

\bibitem[Bonawitz et~al.(2017)Bonawitz, Ivanov, Kreuter, Marcedone, McMahan, Patel, Ramage, Segal, and Seth]{bonawitz2017federated}
Bonawitz, K., Ivanov, V., Kreuter, B., Marcedone, A., McMahan, B., Patel, S., Ramage, D., Segal, A., and Seth, K.
\newblock Practical secure aggregation for privacy-preserving machine learning.
\newblock In \emph{Proceedings of the 2017 ACM SIGSAC Conference on Computer and Communications Security}, pp.\  1175--1191, 2017.

\bibitem[Boneh \& Franklin(1997)Boneh and Franklin]{BFdistributedRSA}
Boneh, D. and Franklin, M.~K.
\newblock Efficient generation of shared rsa keys (extended abstract).
\newblock In \emph{Advances in Cryptology - CRYPTO '97, 17th Annual International Cryptology Conference, Santa Barbara, California, USA, August 17-21, 1997, Proceedings}, volume 1294 of \emph{Lecture Notes in Computer Science}, pp.\  425--439. Springer, 1997.
\newblock \doi{10.1007/BFb0052253}.

\bibitem[Castro \& Liskov(1999)Castro and Liskov]{castro1999practical}
Castro, M. and Liskov, B.
\newblock Practical byzantine fault tolerance.
\newblock In \emph{Proceedings of the Third Symposium on Operating Systems Design and Implementation}, pp.\  173--186, 1999.

\bibitem[Cavoukian(2009)]{cavoukian2009privacy}
Cavoukian, A.
\newblock Privacy by design: The 7 foundational principles.
\newblock \emph{Information and Privacy Commissioner of Ontario}, 2009.

\bibitem[Chen et~al.(2024)Chen, Bandy, Buckley, and Bhatia]{christchurch2024}
Chen, J., Bandy, J., Buckley, D., and Bhatia, R.
\newblock Ai transparency in practice: What was learnt from third-party audit of recommender systems at linkedin and dailymotion.
\newblock \emph{The Christchurch Call}, 2024.
\newblock URL \url{https://www.christchurchcall.org/safe-secure-private-research-finds-third-parties-can-audit-online-algorithms/}.

\bibitem[Cheu et~al.(2019)Cheu, Smith, Ullman, Zeber, and Zhilyaev]{cheu2019distributed}
Cheu, A., Smith, A., Ullman, J., Zeber, D., and Zhilyaev, M.
\newblock Distributed differential privacy via shuffling.
\newblock In \emph{Annual International Conference on the Theory and Applications of Cryptographic Techniques (EUROCRYPT)}, pp.\  375--403, 2019.

\bibitem[Corrigan-Gibbs \& Boneh(2017)Corrigan-Gibbs and Boneh]{corrigan2017prio}
Corrigan-Gibbs, H. and Boneh, D.
\newblock Prio: Private, robust, and scalable computation of aggregate statistics.
\newblock In \emph{14th USENIX Symposium on Networked Systems Design and Implementation (NSDI)}, pp.\  259--282, 2017.

\bibitem[Damg{\aa}rd et~al.(2012)Damg{\aa}rd, Pastro, Smart, and Zakarias]{damgard2012multiparty}
Damg{\aa}rd, I., Pastro, V., Smart, N., and Zakarias, S.
\newblock Multiparty computation from somewhat homomorphic encryption.
\newblock In \emph{Annual International Cryptology Conference}, pp.\  643--662. Springer, 2012.

\bibitem[Dean \& Ghemawat(2008)Dean and Ghemawat]{mapreduce}
Dean, J. and Ghemawat, S.
\newblock Mapreduce: simplified data processing on large clusters.
\newblock \emph{Communications of the ACM}, 51, 2008.
\newblock URL \url{http://doi.acm.org/10.1145/1327452.1327492}.

\bibitem[Dingledine et~al.(2004)Dingledine, Mathewson, and Syverson]{dingledine2004tor}
Dingledine, R., Mathewson, N., and Syverson, P.
\newblock Tor: The second-generation onion router.
\newblock In \emph{13th USENIX Security Symposium}, pp.\  303--320, 2004.

\bibitem[Dorri et~al.(2017)Dorri, Kanhere, Jurdak, and Gauravaram]{dorri2017blockchain}
Dorri, A., Kanhere, S.~S., Jurdak, R., and Gauravaram, P.
\newblock Blockchain for iot security and privacy: The case study of a smart home.
\newblock \emph{IEEE International Conference on Pervasive Computing and Communications Workshops}, pp.\  618--623, 2017.

\bibitem[Duch-Brown et~al.(2017)Duch-Brown, Martens, and Mueller-Langer]{duch2017data}
Duch-Brown, N., Martens, B., and Mueller-Langer, F.
\newblock The economics of ownership, access and trade in digital data.
\newblock \emph{JRC Digital Economy Working Paper}, 2017.

\bibitem[Dwork \& Roth(2014)Dwork and Roth]{dwork2014algorithmic}
Dwork, C. and Roth, A.
\newblock The algorithmic foundations of differential privacy.
\newblock \emph{Foundations and Trends in Theoretical Computer Science}, 9\penalty0 (3--4):\penalty0 211--407, 2014.

\bibitem[Furukawa et~al.(2017)Furukawa, Lindell, Nof, and Weinstein]{FLNW17}
Furukawa, J., Lindell, Y., Nof, A., and Weinstein, O.
\newblock High-throughput secure three-party computation for malicious adversaries and an honest majority.
\newblock In \emph{Advances in Cryptology---EUROCRYPT}, 2017.

\bibitem[Gennaro et~al.(2010)Gennaro, Gentry, and Parno]{gennaro2010non}
Gennaro, R., Gentry, C., and Parno, B.
\newblock Non-interactive verifiable computing: Outsourcing computation to untrusted workers.
\newblock In \emph{Annual International Cryptology Conference}, pp.\  465--482. Springer, 2010.

\bibitem[Gentry(2009{\natexlab{a}})]{FHE}
Gentry, C.
\newblock \emph{A fully homomorphic encryption scheme}.
\newblock PhD thesis, Stanford University, 2009{\natexlab{a}}.
\newblock \url{crypto.stanford.edu/craig}.

\bibitem[Gentry(2009{\natexlab{b}})]{gentry2009fully}
Gentry, C.
\newblock Fully homomorphic encryption using ideal lattices.
\newblock In \emph{ACM Symposium on Theory of Computing (STOC)}, pp.\  169--178, 2009{\natexlab{b}}.

\bibitem[Geoghegan et~al.(2025)Geoghegan, Patton, Pitman, Rescorla, and Wood]{dap-16}
Geoghegan, T., Patton, C., Pitman, B., Rescorla, E., and Wood, C.~A.
\newblock {Distributed Aggregation Protocol for Privacy Preserving Measurement}.
\newblock Internet-Draft draft-ietf-ppm-dap-16, Internet Engineering Task Force, September 2025.
\newblock URL \url{https://datatracker.ietf.org/doc/draft-ietf-ppm-dap/16/}.
\newblock Work in Progress.

\bibitem[Ghorbani \& Zou(2019)Ghorbani and Zou]{ghorbani2019data}
Ghorbani, A. and Zou, J.
\newblock Data shapley: Equitable valuation of data for machine learning.
\newblock In \emph{International Conference on Machine Learning (ICML)}, pp.\  2242--2251, 2019.

\bibitem[Goldwasser \& Kalai(2019)Goldwasser and Kalai]{goldwasser2019cryptographic}
Goldwasser, S. and Kalai, Y.~T.
\newblock Cryptographic assumptions: A position paper.
\newblock \emph{Theory of Computing}, 15\penalty0 (1):\penalty0 1--31, 2019.

\bibitem[He et~al.(2020)He, Li, So, Zeng, Zhang, Wang, Wang, Vepakomma, Singh, Qiu, et~al.]{he2020fedml}
He, C., Li, S., So, J., Zeng, X., Zhang, M., Wang, H., Wang, X., Vepakomma, P., Singh, A., Qiu, H., et~al.
\newblock Fedml: A research library and benchmark for federated machine learning.
\newblock \emph{arXiv preprint arXiv:2007.13518}, 2020.

\bibitem[Hoofnagle et~al.(2019)Hoofnagle, van~der Sloot, and Borgesius]{hoofnagle2019european}
Hoofnagle, C.~J., van~der Sloot, B., and Borgesius, F.~Z.
\newblock The european union general data protection regulation: What it is and what it means.
\newblock \emph{Information \& Communications Technology Law}, 28\penalty0 (1):\penalty0 65--98, 2019.

\bibitem[Ichihashi(2021)]{ichihashi2021economics}
Ichihashi, S.
\newblock The economics of data externalities.
\newblock \emph{Journal of Economic Theory}, 196:\penalty0 105316, 2021.

\bibitem[Jawalkar et~al.(2024)Jawalkar, Gupta, Basu, Chandran, Gupta, and Sharma]{orca}
Jawalkar, N., Gupta, K., Basu, A., Chandran, N., Gupta, D., and Sharma, R.
\newblock Orca: Fss-based secure training and inference with gpus.
\newblock In \emph{2024 IEEE Symposium on Security and Privacy (SP)}, 2024.

\bibitem[Jia et~al.(2019)Jia, Dao, Wang, Hubis, Hynes, G{\"u}rel, Li, Zhang, Song, and Spanos]{jia2019towards}
Jia, R., Dao, D., Wang, B., Hubis, F.~A., Hynes, N., G{\"u}rel, N.~M., Li, B., Zhang, C., Song, D., and Spanos, C.~J.
\newblock Towards efficient data valuation based on the shapley value.
\newblock In \emph{International Conference on Artificial Intelligence and Statistics}, pp.\  1167--1176, 2019.

\bibitem[Jones \& Tonetti(2020)Jones and Tonetti]{jones2020nonrivalry}
Jones, C.~I. and Tonetti, C.
\newblock Nonrivalry and the economics of data.
\newblock \emph{American Economic Review}, 110\penalty0 (9):\penalty0 2819--2858, 2020.

\bibitem[Kairouz et~al.(2021)Kairouz, McMahan, Avent, Bellet, Bennis, Bhagoji, Bonawitz, Charles, Cormode, Cummings, et~al.]{Kairouz_fl}
Kairouz, P., McMahan, H.~B., Avent, B., Bellet, A., Bennis, M., Bhagoji, A.~N., Bonawitz, K., Charles, Z., Cormode, G., Cummings, R., et~al.
\newblock Advances and open problems in federated learning.
\newblock \emph{Foundations and Trends{\textregistered} in Machine Learning}, 14\penalty0 (1--2):\penalty0 1--210, 2021.

\bibitem[Khan \& Vaheesan(2017)Khan and Vaheesan]{khan2017market}
Khan, L.~M. and Vaheesan, S.
\newblock Market power and inequality: The antitrust counterrevolution and its discontents.
\newblock \emph{Harv. L. \& Pol'y Rev.}, 11:\penalty0 235, 2017.

\bibitem[Li et~al.(2020)Li, Sahu, Talwalkar, and Smith]{li2020federated}
Li, T., Sahu, A.~K., Talwalkar, A., and Smith, V.
\newblock Federated learning: Challenges, methods, and future directions.
\newblock \emph{IEEE Signal Processing Magazine}, 37\penalty0 (3):\penalty0 50--60, 2020.

\bibitem[Liu et~al.(2021)Liu, Fan, Chen, Xu, and Yang]{liu2021fate}
Liu, Y., Fan, T., Chen, T., Xu, Q., and Yang, Q.
\newblock Fate: An industrial grade platform for collaborative learning with data protection.
\newblock In \emph{Journal of Machine Learning Research}, volume~22, pp.\  1--6, 2021.

\bibitem[Lycklama et~al.(2024)Lycklama, Viand, Küchler, Knabenhans, and Hithnawi]{auditml}
Lycklama, H., Viand, A., Küchler, N., Knabenhans, C., and Hithnawi, A.
\newblock Holding secrets accountable: Auditing privacy-preserving machine learning, 2024.
\newblock URL \url{https://arxiv.org/abs/2402.15780}.

\bibitem[Martens(2021)]{martens}
Martens, B.
\newblock An economic perspective on data and platform market power.
\newblock \emph{SSRN Electronic Journal}, 02 2021.
\newblock \doi{10.2139/ssrn.3783297}.

\bibitem[Mitrea et~al.(2023)Mitrea, Cioara, and Anghel]{mitrea2023privacy}
Mitrea, D., Cioara, T., and Anghel, I.
\newblock Privacy-preserving computation for peer-to-peer energy trading on a public blockchain.
\newblock \emph{Sensors}, 23\penalty0 (10):\penalty0 4640, 2023.

\bibitem[Mohassel \& Rindal(2018)Mohassel and Rindal]{mohassel2018aby3}
Mohassel, P. and Rindal, P.
\newblock Aby3: A mixed protocol framework for machine learning.
\newblock In \emph{Proceedings of the 2018 ACM SIGSAC Conference on Computer and Communications Security}, pp.\  35--52, 2018.

\bibitem[Moon et~al.(2024)Moon, Yoo, Jiang, and Kim]{miran_thor}
Moon, J., Yoo, D., Jiang, X., and Kim, M.
\newblock {THOR}: Secure transformer inference with homomorphic encryption.
\newblock Cryptology {ePrint} Archive, Paper 2024/1881, 2024.
\newblock URL \url{https://eprint.iacr.org/2024/1881}.

\bibitem[Moore(2025)]{data50}
Moore, E.
\newblock “is \$50 a fair price for your data?”.
\newblock \emph{Financial Times}, September 2025.
\newblock URL \url{https://www.ft.com/content/ea7667fb-8e88-4d1b-9912-8cbfd8b27d43}.
\newblock Accessed: 2025-09-18.

\bibitem[Nagle et~al.(2025)Nagle, Seamans, and Tadelis]{nagle}
Nagle, F., Seamans, R., and Tadelis, S.
\newblock Transaction cost economics in the digital economy: A research agenda.
\newblock \emph{Strategic Organization}, 23\penalty0 (2):\penalty0 351--365, 2025.
\newblock \doi{10.1177/14761270241228674}.
\newblock URL \url{https://doi.org/10.1177/14761270241228674}.

\bibitem[{New Public}()]{newpublic_blacksky}
{New Public}.
\newblock Social media’s next evolution: decentralized, open-source, and scalable.
\newblock \url{https://newpublic.substack.com/p/how-blacksky-grew-to-millions-of}.
\newblock Accessed: 2025-09-25.

\bibitem[{OpenDP Project}(2024)]{opendp_library}
{OpenDP Project}.
\newblock {OpenDP Library: A framework for building differentially private computations}.
\newblock \url{https://opendp.org/tools/}, 2024.
\newblock Accessed: \today.

\bibitem[Posner \& Weyl(2018)Posner and Weyl]{posner2018radical}
Posner, E.~A. and Weyl, E.~G.
\newblock \emph{Radical Markets: Uprooting Capitalism and Democracy for a Just Society}.
\newblock Princeton University Press, 2018.

\bibitem[Rathee et~al.(2022)Rathee, Shen, Wagh, and Popa]{elsa}
Rathee, M., Shen, C., Wagh, S., and Popa, R.~A.
\newblock {ELSA}: Secure aggregation for federated learning with malicious actors.
\newblock Cryptology {ePrint} Archive, Paper 2022/1695, 2022.
\newblock URL \url{https://eprint.iacr.org/2022/1695}.

\bibitem[Roughgarden et~al.(2010)Roughgarden, Tardos, and Vazirani]{roughgarden2010algorithmic}
Roughgarden, T., Tardos, E., and Vazirani, V.~V.
\newblock \emph{Algorithmic Game Theory}.
\newblock Cambridge University Press, 2010.

\bibitem[Ryffel et~al.(2018)Ryffel, Trask, Dahl, Wagner, Mancuso, Rueckert, and Passerat-Palmbach]{ryffel2018generic}
Ryffel, T., Trask, A., Dahl, M., Wagner, B., Mancuso, J., Rueckert, D., and Passerat-Palmbach, J.
\newblock A generic framework for privacy preserving deep learning.
\newblock \emph{arXiv preprint arXiv:1811.04017}, 2018.

\bibitem[Shamir(1979)]{SS}
Shamir, A.
\newblock How to share a secret.
\newblock \emph{Communications of the ACM}, 22\penalty0 (11):\penalty0 612--613, 1979.

\bibitem[Solove(2021)]{solove2021breach}
Solove, D.~J.
\newblock The myth of the privacy paradox.
\newblock \emph{George Washington Law Review}, 89:\penalty0 1--51, 2021.

\bibitem[Statista(2024)]{statista_global_data_2024}
Statista.
\newblock Volume of data/information created, captured, copied, and consumed worldwide from 2010 to 2025 (forecast).
\newblock \url{https://www.statista.com/statistics/871513/worldwide-data-created/}, 2024.

\bibitem[Tamkin et~al.(2024)Tamkin, McCain, Handa, Durmus, Lovitt, Rathi, Huang, Mountfield, Hong, Ritchie, Stern, Clarke, Goldberg, Sumers, Mueller, McEachen, Mitchell, Carter, Clark, Kaplan, and Ganguli]{clio}
Tamkin, A., McCain, M., Handa, K., Durmus, E., Lovitt, L., Rathi, A., Huang, S., Mountfield, A., Hong, J., Ritchie, S., Stern, M., Clarke, B., Goldberg, L., Sumers, T.~R., Mueller, J., McEachen, W., Mitchell, W., Carter, S., Clark, J., Kaplan, J., and Ganguli, D.
\newblock Clio: Privacy-preserving insights into real-world ai use, 2024.
\newblock URL \url{https://arxiv.org/abs/2412.13678}.

\bibitem[Thompson(2024)]{gpt5data}
Thompson, A.~D.
\newblock What’s in gpt-5? a comprehensive analysis of datasets likely used to train gpt-5.
\newblock LifeArchitect.ai, August 2024.
\newblock URL \url{https://lifearchitect.ai/whats-in-gpt-5/}.
\newblock 27 pages, incl.\ title page, references, appendices.

\bibitem[{United Nations Commission on Science and Technology for Development (CSTD)}(2023)]{cstd_data}
{United Nations Commission on Science and Technology for Development (CSTD)}.
\newblock Issues paper on data for development.
\newblock Technical report, {United Nations Conference on Trade and Development (UNCTAD)}, 2023.
\newblock URL \url{https://unctad.org/system/files/information-document/CSTD2023-2024_Issues01_data_en.pdf}.
\newblock Inter-sessional Panel 2023--2024 (Lisbon, Nov. 6--7, 2023); prepared by the UNCTAD Secretariat.

\bibitem[Wagh(2020)]{sameerthesis}
Wagh, S.
\newblock \emph{{New Directions in Efficient Privacy Preserving Machine Learning}}.
\newblock PhD thesis, Princeton University, 2020.

\bibitem[Wagh(2022)]{pika}
Wagh, S.
\newblock Pika: Secure computation using function secret sharing over rings.
\newblock In \emph{Privacy Enhancing Technologies Symposium (PETS)}, 2022.

\bibitem[Wagh et~al.(2019)Wagh, Gupta, and Chandran]{securenn}
Wagh, S., Gupta, D., and Chandran, N.
\newblock {SecureNN: 3-Party Secure Computation for Neural Network Training}.
\newblock In \emph{Privacy Enhancing Technologies Symposium (PETS)}, 2019.

\bibitem[Wagh et~al.(2020)Wagh, He, Machanavajjhala, and Mittal]{dpcrypto}
Wagh, S., He, X., Machanavajjhala, A., and Mittal, P.
\newblock {DP-Cryptography: marrying differential privacy and cryptography in emerging applications}.
\newblock In \emph{Communications of the ACM}, 2020.

\bibitem[Wagh et~al.(2021)Wagh, Tople, Benhamouda, Kushilevitz, Mittal, and Rabin]{falcon}
Wagh, S., Tople, S., Benhamouda, F., Kushilevitz, E., Mittal, P., and Rabin, T.
\newblock {FALCON: Honest-Majority Maliciously Secure Framework for Private Deep Learning}.
\newblock In \emph{Privacy Enhancing Technologies Symposium (PETS)}, 2021.

\bibitem[Wang et~al.(2014)Wang, Tai, Shu, Chang, and Shieh]{fedmapreduce}
Wang, C.-Y., Tai, T.-L., Shu, J.-S., Chang, J.-B., and Shieh, C.-K.
\newblock Federated mapreduce to transparently run applications on multicluster environment.
\newblock In \emph{2014 IEEE International Congress on Big Data}, pp.\  296--303, 2014.
\newblock \doi{10.1109/BigData.Congress.2014.50}.

\bibitem[Wu \& Seneviratne(2025)Wu and Seneviratne]{blockchainIncentives}
Wu, B. and Seneviratne, O.
\newblock Blockchain-based framework for scalable and incentivized federated learning, 2025.
\newblock URL \url{https://arxiv.org/abs/2502.14170}.

\bibitem[Yao et~al.(2024)]{yao2024gac}
Yao, C. et~al.
\newblock Breaking the ceiling of the llm community by treating token generation as a classification for ensembling.
\newblock In \emph{Conference on Empirical Methods in Natural Language Processing (EMNLP)}, 2024.
\newblock URL \url{https://arxiv.org/abs/2406.12585}.

\bibitem[Zheng et~al.(2017)Zheng, Dave, Beekman, Popa, Gonzalez, and Stoica]{zheng2017opaque}
Zheng, W., Dave, A., Beekman, J.~G., Popa, R.~A., Gonzalez, J.~E., and Stoica, I.
\newblock Opaque: An oblivious and encrypted distributed analytics platform.
\newblock In \emph{14th USENIX Symposium on Networked Systems Design and Implementation (NSDI)}, pp.\  283--298, 2017.

\end{thebibliography}
\bibliographystyle{mlsys2025}

\appendix
\section{Complete Experimental Results}\label{app:linked_daily}
Here we describe the main results of the DailyMotion study. Full details of the experiments are provided in this online report~\cite{christchurch2024}. 

\subsection{Dailymotion: Content Moderation and Recommendation Inequality}

\noindent\textbf{Experimental Setup.} Dailymotion provided impression data from 10 million video recommendation records with associated content scoring metadata. The platform deployed PMSR to enable auditors to examine how three distinct recommendation algorithms handled potentially sensitive content, measured through an internal "suggestiveness score" combining violence and sexual content signals. Unlike LinkedIn's differential privacy requirements, Dailymotion opted for policy-based access control with manual approval for each query, demonstrating PMSR's flexibility in accommodating heterogeneous privacy requirements. The same team of four distributed auditors analyzed patterns of content promotion and recommendation inequality across the platform's algorithmic systems.

\textbf{Private Map and Secure Reduce Implementation.} Light Nodes executed queries against video impression logs, extracting the suggestiveness scores, view counts, and algorithm identifiers. The platform provided mock data matching the production schema to enable iterative refinement of analysis code before execution on real data. The study focused on two key questions: whether different algorithms promoted suggestive content at different rates, and how recommendation patterns affected content distribution equality. Each query underwent manual review by Dailymotion's data governance team before approval. Similar to LinkedIn, Secure Reduce was implemented using a single Heavy Node, as the combination of access control and manual approval provided sufficient security guarantees for this use case.

\begin{figure*}[th]
\centering
\includegraphics[width=\textwidth]{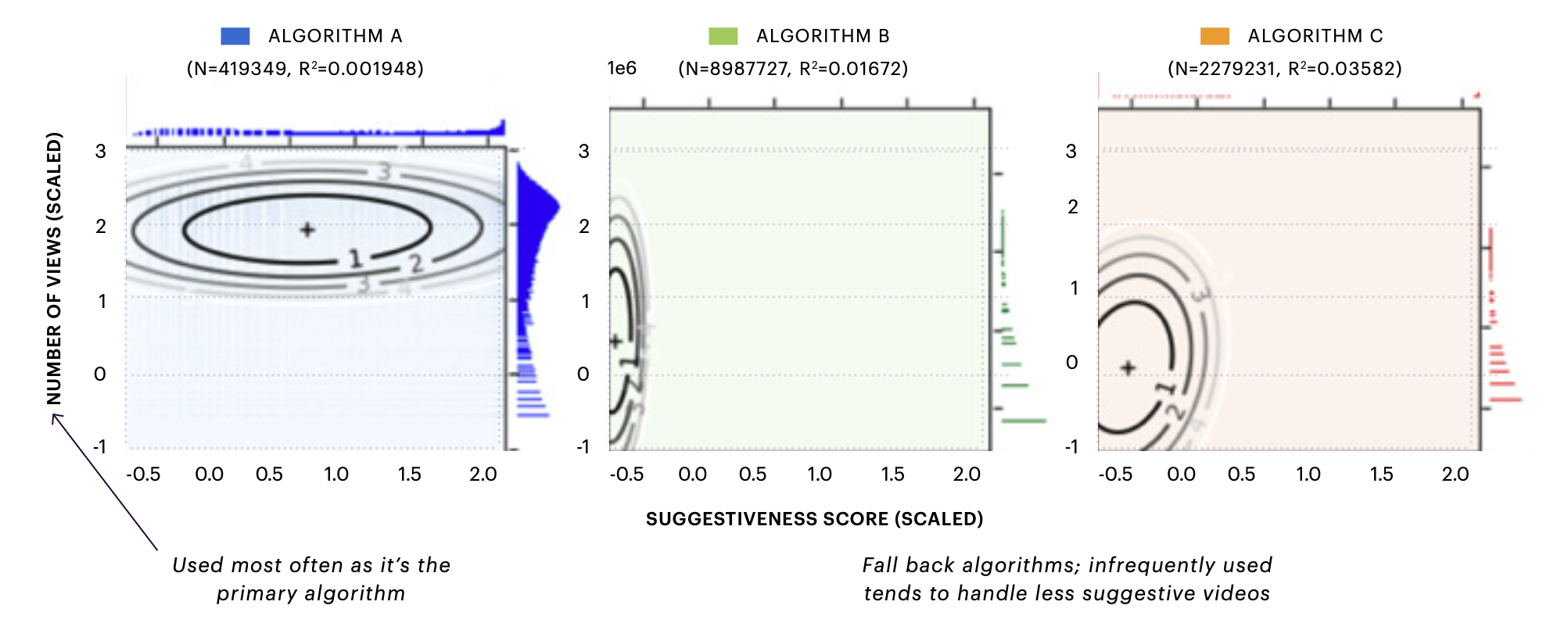}
\caption{Distribution of video suggestiveness scores by recommendation algorithm. Contour plots show density of (suggestiveness, view count) pairs. Algorithms B and C show sharp cutoffs at low suggestiveness values, while Algorithm A operates across the full range.}
\label{fig:suggestiveness}
\end{figure*}

\textbf{Results and Impact.} The audit revealed critical insights into algorithmic content segregation that would have been impossible to discover without access to production data. Figure~\ref{fig:suggestiveness} shows striking differences in content handling: Algorithm A (primary recommender, 77\% of traffic) promoted videos across the full suggestiveness spectrum, while Algorithms B and C exclusively handled content below 0.5 suggestiveness scores, suggesting intentional content segregation for brand safety or regulatory compliance. Additionally, analysis of recommendation inequality using differentially private Gini coefficient computation revealed extreme concentration in Dailymotion's primary algorithm (Gini = 0.68), with the top 10\% of videos capturing 75\% of all recommendations-substantially higher than LinkedIn's more equitable distribution (Gini = 0.31). These findings demonstrate PMSR's capability to uncover both content moderation practices and distributional effects of recommendation algorithms, providing transparency into systems that significantly shape online information ecosystems while maintaining platform confidentiality and user privacy.

\begin{table}[h]
\centering
\caption{Recommendation inequality metrics across platforms}
\begin{tabular}{lcc}
\toprule
Platform & Gini Coefficient & Top 10\% Share \\
\midrule
LinkedIn (AI ranker) & 0.31 ± 0.02 & 25\% \\
LinkedIn (Traditional) & 0.28 ± 0.02 & 23\% \\
Dailymotion (Algo A) & 0.68 ± 0.03 & 75\% \\
Dailymotion (Algo B) & 0.44 ± 0.04 & 48\% \\
\bottomrule
\end{tabular}
\end{table}

\subsection{Detailed Analysis of Distributed AI Inference}

\textbf{Subject-Specific Performance Patterns.} The table below shows the subject wise ensemble, theoretical maximum accuracy, and the number of questions.

\onecolumn 
\begin{longtable}{l c c c c}
\caption{Theoretical Maximum Accuracy Report}
\label{tab:accuracy_report} \\
\toprule
\textbf{Subject} & \textbf{Ensemble} & \textbf{Theo Max} & \textbf{Gain} & \textbf{Questions} \\
\midrule
\endfirsthead

\multicolumn{5}{c}{\tablename\ \thetable{} -- continued from previous page} \\
\toprule
\textbf{Subject} & \textbf{Ensemble} & \textbf{Theo Max} & \textbf{Gain} & \textbf{Questions} \\
\midrule
\endhead

\midrule
\multicolumn{5}{r}{Continued on next page} \\
\endfoot

\bottomrule
\endlastfoot

high\_school\_government\_and\_politics & 0.9896 & 1.0000 & 0.0104 & 193 \\
medical\_genetics & 0.9700 & 1.0000 & 0.0300 & 100 \\
professional\_medicine & 0.9522 & 0.9963 & 0.0441 & 272 \\
astronomy & 0.9539 & 0.9934 & 0.0395 & 152 \\
marketing & 0.9530 & 0.9915 & 0.0385 & 234 \\
management & 0.9223 & 0.9903 & 0.0680 & 103 \\
high\_school\_us\_history & 0.9608 & 0.9902 & 0.0294 & 204 \\
sociology & 0.9552 & 0.9900 & 0.0348 & 201 \\
college\_computer\_science & 0.8300 & 0.9900 & 0.1600 & 100 \\
high\_school\_computer\_science & 0.9400 & 0.9900 & 0.0500 & 100 \\
high\_school\_psychology & 0.9706 & 0.9890 & 0.0183 & 545 \\
high\_school\_microeconomics & 0.9706 & 0.9874 & 0.0168 & 238 \\
elementary\_mathematics & 0.9312 & 0.9868 & 0.0556 & 378 \\
college\_biology & 0.9514 & 0.9861 & 0.0347 & 144 \\
miscellaneous & 0.9566 & 0.9860 & 0.0294 & 783 \\
human\_sexuality & 0.9389 & 0.9847 & 0.0458 & 131 \\
international\_law & 0.9504 & 0.9835 & 0.0331 & 121 \\
conceptual\_physics & 0.9319 & 0.9830 & 0.0511 & 235 \\
clinical\_knowledge & 0.9132 & 0.9811 & 0.0679 & 265 \\
high\_school\_biology & 0.9581 & 0.9806 & 0.0226 & 310 \\
high\_school\_world\_history & 0.9451 & 0.9789 & 0.0338 & 237 \\
moral\_scenarios & 0.8458 & 0.9777 & 0.1318 & 895 \\
prehistory & 0.9024 & 0.9756 & 0.0732 & 41 \\
logical\_fallacies & 0.9141 & 0.9755 & 0.0613 & 163 \\
high\_school\_geography & 0.9545 & 0.9747 & 0.0202 & 198 \\
nutrition & 0.8987 & 0.9739 & 0.0752 & 306 \\
us\_foreign\_policy & 0.9500 & 0.9700 & 0.0200 & 100 \\
high\_school\_macroeconomics & 0.9077 & 0.9692 & 0.0615 & 390 \\
professional\_psychology & 0.9100 & 0.9656 & 0.0556 & 611 \\
college\_physics & 0.8137 & 0.9608 & 0.1471 & 102 \\
high\_school\_physics & 0.8212 & 0.9603 & 0.1391 & 151 \\
computer\_security & 0.8500 & 0.9600 & 0.1100 & 100 \\
world\_religions & 0.9123 & 0.9591 & 0.0468 & 171 \\
electrical\_engineering & 0.8690 & 0.9586 & 0.0897 & 145 \\
philosophy & 0.9132 & 0.9582 & 0.0450 & 311 \\
anatomy & 0.8444 & 0.9556 & 0.1111 & 135 \\
professional\_accounting & 0.8156 & 0.9539 & 0.1383 & 282 \\
jurisprudence & 0.9074 & 0.9537 & 0.0463 & 108 \\
high\_school\_statistics & 0.8657 & 0.9537 & 0.0880 & 216 \\
security\_studies & 0.8449 & 0.9510 & 0.1061 & 245 \\
high\_school\_chemistry & 0.8621 & 0.9507 & 0.0887 & 203 \\
high\_school\_mathematics & 0.7222 & 0.9481 & 0.2259 & 270 \\
high\_school\_european\_history & 0.9091 & 0.9455 & 0.0364 & 165 \\
formal\_logic & 0.7778 & 0.9444 & 0.1667 & 126 \\
college\_medicine & 0.8613 & 0.9422 & 0.0809 & 173 \\
moral\_disputes & 0.8721 & 0.9419 & 0.0698 & 344 \\
human\_aging & 0.8655 & 0.9417 & 0.0762 & 223 \\
abstract\_algebra & 0.7700 & 0.9300 & 0.1600 & 100 \\
global\_facts & 0.6700 & 0.9300 & 0.2600 & 100 \\
machine\_learning & 0.8036 & 0.9286 & 0.1250 & 112 \\
business\_ethics & 0.8600 & 0.9200 & 0.0600 & 100 \\
college\_mathematics & 0.7000 & 0.9200 & 0.2200 & 100 \\
professional\_law & 0.7425 & 0.9146 & 0.1721 & 1534 \\
econometrics & 0.7544 & 0.9123 & 0.1579 & 114 \\
public\_relations & 0.8364 & 0.9091 & 0.0727 & 110 \\
college\_chemistry & 0.6400 & 0.8400 & 0.2000 & 100 \\
virology & 0.6145 & 0.6627 & 0.0482 & 166 \\
\midrule
\textbf{OVERALL} & \textbf{0.8747} & \textbf{0.9589} & \textbf{0.0842} & \textbf{13756} \\
\end{longtable}
\twocolumn 

\end{document}